\documentclass[11pt]{article}
\usepackage{amsmath,amssymb,cite,color,ifpdf}
\usepackage[pdftex]{graphicx}
\usepackage{pifont}
\usepackage{array}
\newcommand{\tick}{{\color{blue}\ding{52}}}
\newcommand{\cross}{{\color{red}\ding{55}}}
\newcommand{\BY}[1]{#1,~}
\newcommand{\IN}[4]{#1~{\bf #2}~(#3)~#4}
\newcommand{\SAME}[3]{~{\bf #1}~(#2)~#3}

\begin{document}

%\vspace{-2.0cm}
\begin{flushright}
Edinburgh 2015/15\\
CERN-PH-TH/2015-177\\
\end{flushright}
\vspace{3cm}

\begin{center}
{\Large \bf Global Parton Distributions for the LHC Run II }
\vspace{1cm}

Richard~D.~Ball

\vspace{1cm}
{\it The Higgs Centre for Theoretical Physics, University of Edinburgh,\\
JCMB, KB, Mayfield Rd, Edinburgh EH9 3JZ, Scotland\\
PH Department, TH Unit, CERN, CH-1211 Geneva 23, Switzerland\\
}
\end{center}

\vspace{3cm}

\begin{center}
 {\bf \large Abstract}
\end{center}
We review the next generation global PDF sets: NNPDF3.0, MMHT14 and CT14. We describe the global datasets, particularly the new data from LHC Run 1, recent developments in QCD theory and PDF methodology, improvements in their combination and delivery, and future prospects for parton determination at Run 2. 

\vspace{2cm}

\noindent Talk given at La Thuile, 3rd March 2015.

\clearpage

\section{Next Generation PDFs}

In order to make the most of the LHC, we need to be able to compute standard model cross-sections reliably and precisely. These days a wide variety of inclusive hard processes are known to NLO and increasingly NNLO in perturbative QCD. However to obtain a physical cross-section, these must be folded with nonperturbative parton distribution functions (PDFs). 
Since the PDFs cannot be computed from first principles, they must be determined empirically. This is a nontrivial task: the PDFs $g$, $u$, $\bar{u}$, $d$, $\bar{d}$, $s$, 
$\bar{s},~\ldots,$ are functions of $x$ and $Q^2$, correlated through both theoretical constraints and measurements from a wide variety of different experiments and processes. Uncertainties in PDFs remain one of the dominant sources of uncertainty for many important LHC cross-sections. Recently, the major PDF collaborations have all been using data from LHC Run I to further constrain PDFs in preparation for Run II. 

There are at present three PDF fitting collaborations providing global PDF determinations. Their most recent sets are NNPDF3.0\cite{ref:NNPDF3} (which supercedes the NNPDF2.x sets\cite{ref:NNPDF2}), MMHT14\cite{ref:MMHT14} (which now replaces the long serving MSTW08 set\cite{ref:MSTW08}), and CT14\cite{ref:CT14} (which supercedes the CT10 sets\cite{ref:CT10}). All three combine a wide range of older DIS, neutrino and Drell-Yan fixed target data with HERA DIS data, Tevatron Drell-Yan, W/Z and jet data, and now also Drell-Yan, W/Z and jet data from LHC Run I. These data span a kinematic range of more than four orders of magnitude in $x$ and six orders of magnitude in $Q^2$, and the wide range of different processes are together sufficient to extract all PDF combinations without theoretical assumptions beyond those embodied in fixed order perturbative QCD. By contrast the ABM sets \cite{ref:ABM} are based only on DIS and Drell-Yan data, with no data from the Tevatron, and have difficulties extrapolating up to LHC energies, while the HERA PDFs \cite{ref:HERA1,ref:HERA2} use only HERA data, and consequently have larger uncertainties than the global sets\cite{ref:bench}. In this short review we thus only consider in detail the three most recent global sets.

\section{Global Datasets}

A detailed comparison of the datasets used in each of the three most recent global fits is presented in Tab.\ref{tab:compardat}, together with the total number of datapoints used. The most striking feature of the table is that while the three collaborations have different detailed preferences, the global datasets are broadly similar in scope and coverage. Thus while CT14 does not use the recent CHORUS $\nu$-DIS data, it retains the older CDHSW and CCFR data. While all three collaborations now use the combined HERA-I data, only NNPDF3.0 also uses HERA-II data.\footnote{The combined HERA-II data have only been published very recently \cite{ref:HERA2}, and will no doubt be incorporated in due time. Preliminary analyses by MMHT and NNPDF suggest that their impact will be small.} NNPDF prefer not to use D0 jet data, which were analysed with the midpoint algorithm which is infrared unsafe and thus cannot be used with NNLO calculations: all three collaborations now use a significant amount of LHC Run I data, though CT14 has yet to include the CMS double differential Drell-Yan data or the $t\bar{t}$ total cross-section. And so on.

\begin{table}[t]
 \centering
{\scriptsize
  \begin{tabular*}{0.61\textwidth}{l||c|c|c|}
    \hline
    & NNPDF3.0 & MMHT14 & CT14(prel)  \\
    \hline
    { SLAC p,d DIS} & \tick & \tick & \cross  \\
     { BCDMS p,d DIS} & \tick & \tick & \tick  \\   
    { NMC p,d DIS} & \tick & \tick & \tick  \\  
     { E665 p,d DIS} & \cross & \tick & \cross  \\  
  { CDHSW nu-DIS} & \cross & \cross & \tick  \\ 
 { CCFR nu-DIS} & \cross & \tick & \tick  \\ 
   { CHORUS nu-DIS} & \tick & \tick & \cross  \\
   { CCFR dimuon} & \cross & \tick & \tick  \\ 
   { NuTeV dimuon} & \tick & \tick & \tick  \\ 
      \hline
{ HERA I NC,CC} & \tick & \tick & \tick  \\
{ HERA I charm} & \tick & \tick & \tick  \\
{ H1,ZEUS jets} & \cross& \tick & \cross  \\
{ H1 HERA II} & \tick & \cross & \cross  \\
{ ZEUS HERA II} & \tick & \cross & \cross \\
    \hline\hline
    { E605 \& E866 FT DY} & \tick & \tick & \tick  \\
    \hline
    { CDF \& D0 W asym} & \cross & \tick & \tick  \\
    { D0 Run II W asym} & \cross & \cross & \tick  \\
   { CDF \& D0 Z rap} & \tick & \tick & \tick  \\
    { CDF Run-II jets} & \tick & \tick & \tick  \\
      { D0 Run-II jets} & \cross & \tick & \tick  \\
\hline 
    { ATLAS high-mass DY} & \tick & \tick & \tick  \\
    { CMS 2D DY} & \tick & \tick & \cross  \\   
    { ATLAS W,Z rap} & \tick & \tick & \tick  \\
    { ATLAS W pT} & \tick & \tick & \cross  \\ 
     { CMS W asy} & \tick & \tick & \tick  \\ 
     { CMS W+c } & \tick & \cross & \cross  \\ 
     { LHCb W,Z rap } & \tick & \tick & \tick  \\
    { ATLAS jets} & \tick & \tick & \tick  \\   
    { CMS jets} & \tick & \tick & \tick  \\
    { ttbar tot xsec} & \tick & \tick & \cross  \\ 
\hline\hline
    {Total NLO} & 4276 & 2996 & 2947 \\ 
    {Total NNLO} & 4078 & 2663 & 2947 \\ 
\hline
  \end{tabular*}}
  \caption{Data included in the latest NLO and NNLO global PDF sets, and the total number of data points in each fit.}
  \label{tab:compardat}
\end{table}

It is expected that over the next few years many more LHC datasets will be added to this list, some of them improvements on existing measurements, others more novel \cite{ref:PDF4LHC}. Light flavour separation will be improved by differential high and low mass Drell-Yan, and more accurate W/Z asymmetries and rapidity distributions, while better W+c data will pin down strangeness, and Z+c and Z+b will assist the direct determination of heavy quark distributions. The gluon at medium and large x will be further constrained by differential top production, inclusive jets and dijets, prompt photons, and W/Z + jets. 

All three collaborations producing global fits now make full use of experimental systematics when implementing new datasets. These systematics can be either additive or multiplicative: multiplicative systematic uncertainties need careful treatment in order to avoid the well known d'Agostini bias \cite{ref:norm}.

\section{Theory and Methodology}

\subsection{Theory}

Each of the three collaborations now produces families of fits at LO (for Monte Carlos), NLO and NNLO in perturbative QCD. All now fit the strangeness distribution $s+\bar{s}$, and NNPDF and MMHT also attempt to fit the strange valence $s-\bar{s}$. None of the currently available sets include fitted charm distributions, though there have been recent studies by CTEQ\cite{ref:CTIC}. All three collaborations use a GM-VFNS for heavy quark distributions (FONLL for NNPDF3.0, TR$^\prime$ for MMHT14 and S-ACOT for CT14, differing only by subleading terms \cite{ref:LHbench}): this is essential for accurate extrapolation to the high scales of many LHC measurements \cite{ref:bench,ref:NNPDFth}. The PDF sets are each determined using $\alpha_s(m_Z)=0.118$, but provide other sets with $\alpha_s$ either side of this value (at intervals of $0.001$) for determination of $\alpha_s$ uncertainties. They also have their own preferred values of $\alpha_s$ (at NNLO these are $0.1173\pm 0.0007$ \cite{ref:NNPDFas}, $0.1172\pm 0.0013$ \cite{ref:MMHTas} and $0.115^{+0.006}_{-0.004}$ \cite{ref:CT14}). There is as yet no consensus on the input values of $m_c$ and $m_b$, or on whether to use $\overline{\rm MS}$ or pole masses.

An important limitation on the usefulness of hadronic data in constraining PDFs is the availability of NNLO corrections. The recent calculation of the $t\bar{t}$ total cross-section to NNLO \cite{ref:NNLOtop} has had a significant impact on the determination of the gluon distribution, which is expected to improve further once more differential results become available. Calculations of the inclusive jet cross-section to NNLO are now available in the $gg$ and $qq$ channels \cite{ref:NNLOjet}, and the full result is eagerly awaited.

Computationally, new interface tools such as FastNLO \cite{ref:FastNLO} and APPLGRID \cite{ref:APPLGRID} have been developed to evaluate hadronic cross-sections sufficiently fast to be usable in PDF fits. These work by precomputing hard cross-sections in lookup tables. Other tools released recently include a PDF plotting tool APFEL \cite{ref:APFEL} and a general purpose fitting tool HERAfitter \cite{ref:HERAfitter}. The impact of new datasets on PDFs may be estimated using Bayesian reweighting \cite{ref:RW} or PDF profiling \cite{ref:PDF4LHC} as implemented in HERAfitter.

\subsection{Methodology}

Considerable progress has been made over the years in the methodology used to determine PDFs and their uncertainties (see Tab.\ref{tab:comparmeth}). The Hessian method adopts a fixed parametrization, with uncertainties determined through diagonalization of the Hessian matrix. As data become more precise, the parametrization must be more flexible, and MMHT14 and CT14 have recently introduced Chebyshev and Bernstein polynomials into their parametrizations for this purpose. If $\Delta \chi^2=1$ is used to determine uncertainties in this method, PDF errors turn out to be unrealistically small: consequently both collaborations use a tolerance criterion, in which uncertainties are inflated dynamically for each eigenvector in turn in order to maintain errors consistent with those of the data. There has been much speculation as to whether tolerance is required because of defects of the methodology (in particular the limitations of a fixed parametrization), or whether it is due to data inconsistencies or defects of the theoretical tools (in particular fixed order perturbative QCD) used to describe it \cite{ref:pumplin}.

\begin{table}[t!]
  \centering{\scriptsize
  \begin{tabular*}{0.89\textwidth}{l||c|c|c|}
    \hline
    &  NNPDF3.0 &  MMHT14 & CT14  \\
    \hline
    {No.~of fitted PDFs} & 7 & 7 & 6 \\
    {Parametrization} & $x^a(1-x)^b\times$ neural nets & $x^a(1-x)^b\times$ Chebyshev & $x^a(1-x)^b\times$Bernstein \\
    {Free parameters} & 259 & 37 & 28 \\
    {Uncertainties} & MC Replicas & Hessian & Hessian \\
    {Tolerance} & None & Dynamical & Dynamical  \\
    {Closure test} & \tick & \cross & \cross  \\
    {Reweighting} & replicas & eigenvectors & eigenvectors  \\
    \hline
  \end{tabular*}}
  \caption{Main methodological features of various global PDF sets.}
  \label{tab:comparmeth}
\end{table}

The NNPDF collaboration uses instead a Monte Carlo method \cite{ref:GK} in which fits are made to data replicas using a very redundant parametrization (for which NNPDF use a neural network). These fits give an ensemble of PDF replicas, each of which is equally probable, and may thus be used to determine central values, uncertainties, correlations, etc. There is no assumption in this method that the PDF uncertainties are Gaussian. Moreover since there is no $\Delta \chi^2$ criterion, there is no need for tolerance. The redundancy of the parametrization ensures freedom from parametrization bias. 

\begin{figure}[t!]
\begin{center}
\includegraphics[width=0.65\textwidth]{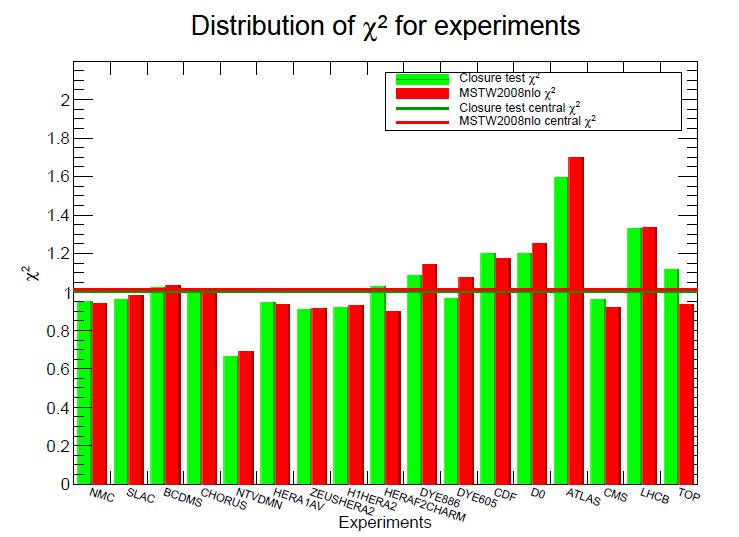}
\vskip 1cm   
\includegraphics[width=0.65\textwidth]{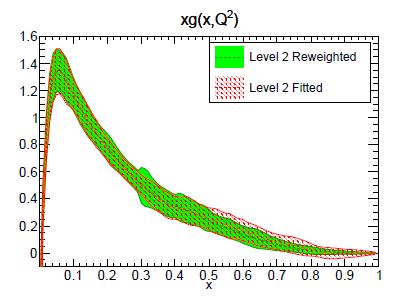} 
\caption{Some results from a closure test: $\chi^2$ values for different datasets (upper) and a reweighting test of the gluon distribution (lower) \cite{ref:NNPDF3}.}
\label{fig:closure}
\end{center}
\vskip-0.5cm
\end{figure}

\begin{figure}[t!]
\begin{center} 
\includegraphics[width=0.6\textwidth]{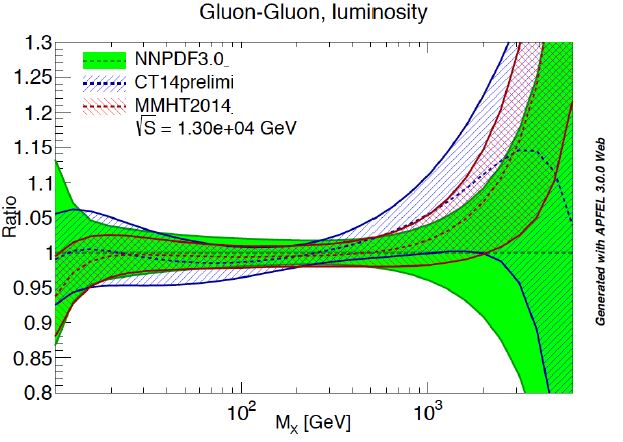} 
\includegraphics[width=0.6\textwidth]{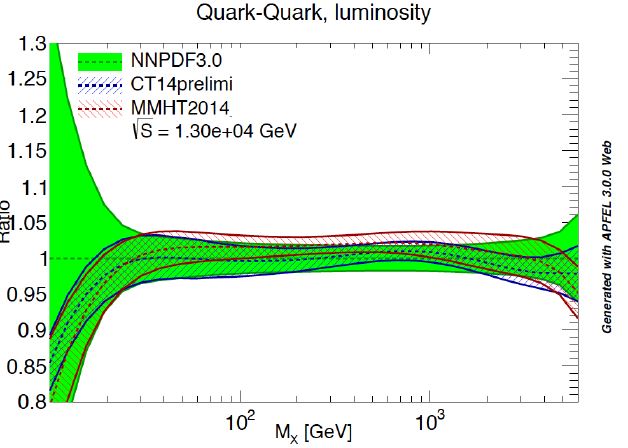}   
\includegraphics[width=0.6\textwidth]{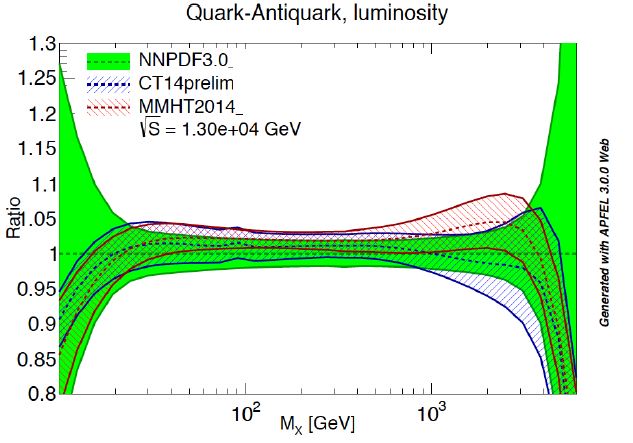} 
\caption{The $gg$, $qq$ and $q\bar{q}$ luminosities (top to bottom) at the LHC with centre of mass energy $13$ TeV, as predicted by the three global PDF sets NNPDF3.0, MMHT14, CT14, normalised to NNPDF3.0}
\label{fig:lumis}
\end{center}
\vskip-0.5cm
\end{figure}

The NNPDF methodology has recently been subjected to a closure test \cite{ref:NNPDF3}. The idea behind this is that if both the data and the theory used to describe them were `perfect', and thus free from any inconsistencies, a fit to these data should also be perfect: any defects in the PDFs would be due entirely to imperfections in the methodology. So in a closure test we assume a given theory (eg NLO QCD), a given prior PDF set $f_0$ (eg MSTW08), and then generate a set of $N$ pseudodata by Monte Carlo, using the assumed theory, $f_0$, and the statistical and systematic uncertainties from a typical global dataset (to ensure the test is as realistic as possible). These perfect pseudodata, together with their uncertainties, are then fitted, to yield a fitted PDF set $f$: if the fitting methodology were perfect, we would then find that $\chi^2=N$, and $f=f_0$, within the PDF uncertainties determined in the fit.

Results from a typical closure test are shown in Fig.\ref{fig:closure}: current NNPDF methodology passes the closure test, in the sense that methodological uncertainties have been demonstrated to be considerably smaller than data and theory uncertainties. This means that uncertainties in NNPDF fits are true statistical uncertainties: the NNPDF probability distributions are a genuine consequence of the prior data and theory that goes into the fit. It would be interesting to also subject the Hessian method to a closure test: in this way it may be possible to understand better the reason for the need for dynamical tolerance, and whether there is any residual bias in central values due to the fixed parametrization.

\section{Results}

For descriptions and plots of the latest global PDFs, and the quality of their description of the various datasets, we refer the reader to the original publications \cite{ref:NNPDF3,ref:MMHT14,ref:CT14}. Here we discuss two subjects of particular interest: the predictions for parton luminosities at $13$ TeV, the strangeness fraction, and recent progress in combination and delivery.

\begin{figure}[t!]
\begin{center}
\includegraphics[width=0.60\textwidth]{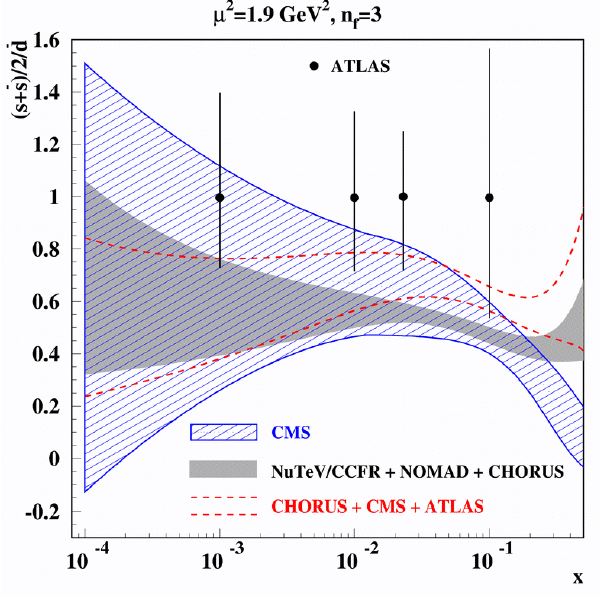}   
\includegraphics[width=0.65\textwidth]{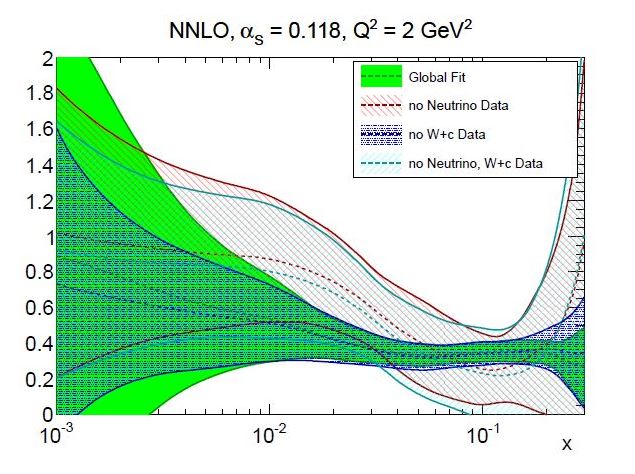} 
\caption{The strangeness fraction $r_s$: results by ATLAS, CMS and neutrino experiments NuTev, NOMAD and CHORUS (above) and from an NNPDF study in the context of a global fit (below). Note that the definitions of $r_s$ used in the two plots are slightly different: in the upper plot $r_s = (s + \bar{s})/2\bar{d}$, while in the lower plot $r_s = (s + \bar{s})/(\bar{u}+\bar{d})$}
\label{fig:strange}
\end{center}
\vskip-0.5cm
\end{figure}

\subsection{Luminosities}

Predictions for the $gg$, $qq$ and $q\bar{q}$ luminosities at the LHC with centre of mass energy $13$ TeV are shown in Fig.\ref{fig:lumis}. In the central region all three collaborations now make consistent predictions, with similar uncertainties: this is particularly noticeable in the $gg$ channel, of direct relevance to Higgs production through gluon fusion, and top production. The $qq$ and $q\bar{q}$ luminosities are also in broad agreement in the central region, but at high scales NNPDF lies above the others, with a substantially larger uncertainty. This is because PDFs at large $x$ are largely unconstrained by data, but must be bounded below by the positivity of any physical cross-section: the uncertainties are thus asymmetrical, and liable to be underestimated by Hessian treatments. Constraints on luminosities at high invariant mass are important for putting bounds on new physics, and deserve more careful study \cite{ref:CJ12}. 

\subsection{Strangeness}

There has been some controversy recently about the strangeness fraction
$r_s(x,Q^2)$ 
with results from ATLAS W+c data apparently suggesting $r_s=1$, albeit with large uncertainties. If confirmed this would overturn conventional wisdom that strangeness should be suppressed due to the strange quark mass. However CMS do see a suppression at large $x$, and this is supported by a recent analysis of neutrino data \cite{ref:strange} (see Fig.\ref{fig:strange}). All the global PDF determinations see strangeness suppression, and a detailed study in the context of the global fits shows that there is little or no tension between the neutrino data and W+c data. It will be interesting to see how this situation develops when we have more precise W+c data from Run 2.

\subsection{Combination and Delivery}

For many years now the PDF4LHC recommendation for combining predictions obtained with different PDF sets was to compute with each of the three global sets \cite{ref:NNPDF2,ref:MSTW08,ref:CT10}, and take the envelope of the resulting predictions \cite{ref:envelope}. This is a conservative method, appropriate for the older PDF sets which displayed some inconsistencies, most noticeably for Higgs production. It was also time consuming.

\begin{figure}[t!]
\begin{center}
\includegraphics[width=0.65\textwidth]{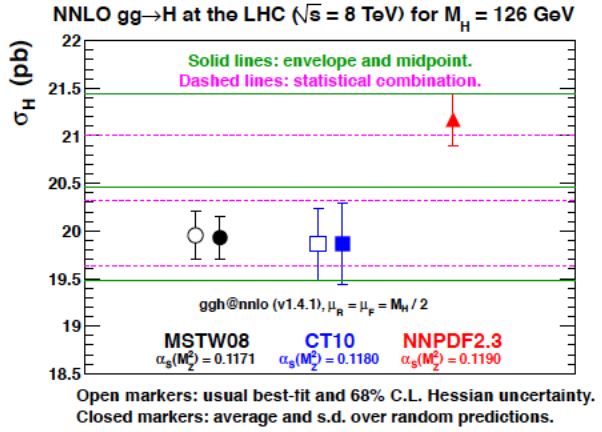} 
\vskip0.5cm
\includegraphics[width=0.65\textwidth]{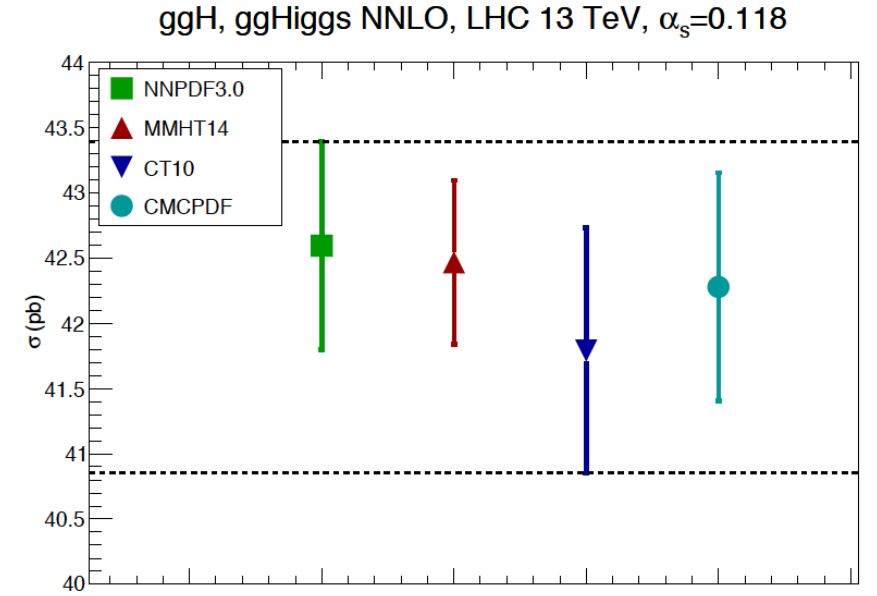} 
\caption{The Higgs gluon fusion cross-section, computed using the old envelope method (upper) and the new combination method (lower).}
\label{fig:ggH}
\end{center}
\vskip-0.5cm
\end{figure}

Since the latest global PDFs are much more consistent between each other, it now becomes possible to combine them statistically into a single PDF set (to be called PDF4LHC15), which becomes the basis for a new recommendation. The combination is done by generating 300 replicas for each PDF set (the replicas for the Hessian sets being produced by a code developed by Thorne and Watt \cite{ref:reps}), to give a set of 900 replicas: the prior assumption in the combination is thus that the global PDFs are not statistically independent, but that each global set is equally probable. The combination is performed at a fixed value of $\alpha_s=0.118$: the $\alpha_s$ uncertainty is treated independently of the PDF uncertainty, and added in quadrature. The results of the old and new procedures for the Higgs gluon fusion NNLO cross-section are shown in Fig.\ref{fig:ggH}.

Since delivery of the full set of 900 replicas is impractical, a number of techniques have been developed to make the combined set more manageable. A replica compression technique, which preserves the non-Gaussian features of the underlying probability distribution, reduces the set of 900 to 100 or less, with little loss of accuracy \cite{ref:compression}. However for many purposes a Hessian representation is preferred, particularly when PDF uncertainties are to be treated as nuisance parameters. To turn replicas into Hessian two approaches have been proposed. The Meta-PDF approach refits to a functional form at a particular scale, which is then evolved in the usual way \cite{ref:metapdf}. The MC2hessian approach instead uses the replicas themselves as a basis set, optimised using a genetic algorithm \cite{ref:MC2H}: in this way no evolution is required since each replica itself contains its own evolution. It is expected that the PDF4LHC15 set will be delivered in three representations: 
\begin{itemize}
\item a small Hessian set with only 30 eigenvectors (for applications where high precision is not required, such as acceptances, efficiencies or extrapolations); 
\item a larger Hessian set with 100 eigenvectors (for PDF uncertainties in precision calculations); and 
\item a Monte Carlo set of 100 replicas (for applications where non-Gaussianity may be important, for example searches). 
\end{itemize}
There will be additional eigenvectors and replicas to allow for $\alpha_s$ variations, the results to be added in quadrature.

\section{Future Prospects}

The determination of global PDFs has made significant advances in recent years: in the inclusion of new and better data (in particularly from LHC), in theoretical advances (driven particularly by new NNLO calculations, and new computational tools), in methodological developments (more flexible parametrizations, closure testing, reweighting and profiling), and in presentational improvements (combination and compression). No doubt many of these lines of development will continue, stimulated by improved data from Run 2.

\subsection{Variations}

Meanwhile, alongside the mainstream work, there are various side projects aimed at broadening the scope and applicability of PDF determination. Electroweak corrections can make substantial contributions to a number of important hadronic processes, particularly W/Z production and top production.  However a consistent calculation of these effects require PDFs with QED corrections, in particular with a fitted photon PDF. A first global determination of the photon PDF and its uncertainties, using LO QED and NNLO QCD, was performed recently \cite{ref:NNPDFQED}, but uncertainties are still large. The situation may improve in the future following a more detailed study of processes such as $W$ pair production which may further constrain the photon PDF.

Fixed order perturbative QCD becomes increasingly unreliable at large $x$ and small $x$ due to unresummed logarithms. Evidence for the effect of small $x$ (high energy) logarithms has been reported by the HERA collaboration \cite{ref:HERA1,ref:HERA2}, but as yet there are no global fits which include the effects of small $x$ resummation. However a global fit which resums large $x$ (or threshold) logarithms was performed recently \cite{ref:NNPDFres}, and will have implications for searches for new physics since the resummation significantly reduces the quark luminosities at high invariant mass. Uncertainties are still large, however. 

All three global PDF collaborations attempt to exclude higher twist and (to some extent) nuclear effects by cutting fixed target data at low $Q^2$ and $W^2$. These cuts are generally effective \cite{ref:NNPDFth}. Various attempts have been made to model higher twist and nuclear effects \cite{ref:CJ12,ref:nCTEQ,ref:MMHT14}, one of the aims being to improve the accuracy of PDFs in the large $x$ region by a controlled relaxation of the $W^2$ cut. An alternative strategy would be to eliminate the use of fixed target data altogether, but the uncertainties on collider-only fits \cite{ref:NNPDF3,ref:NNPDF2,ref:HERA1,ref:HERA2} are still too great for them to be competitive with the global fits.

A first global determination of spin dependent PDFs and their uncertainties was also performed recently \cite{ref:NNPDFpol}, supplementing polarized DIS data with polarized inclusive jet and W production data from RHIC. While there is some evidence for a polarized gluon distribution at large $x$, first moments remain elusive due to the limited small $x$ reach of the data.

\subsection{Theory Uncertainties}

\begin{figure}[t!]
\begin{center}
\includegraphics[width=0.65\textwidth]{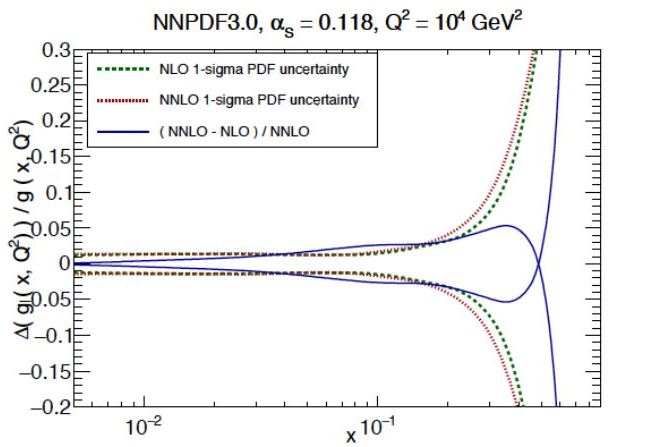} 
%\vskip 1cm 
\includegraphics[width=0.65\textwidth]{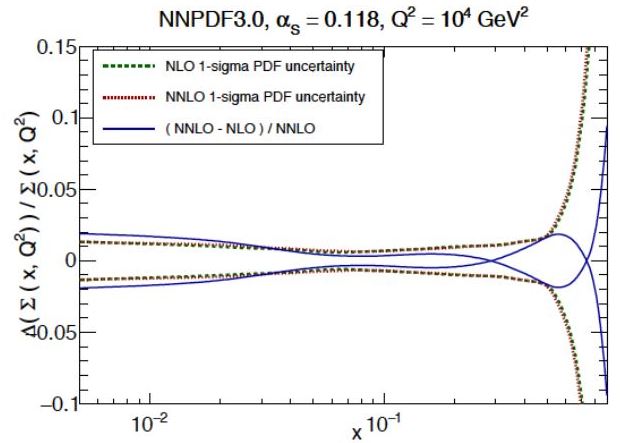} 
\caption{An estimate of the theory uncertainty due to higher order corrections in the NLO gluon, obtained by comparing the result of NLO and NNLO fits.}
\label{fig:th}
\end{center}
\vskip-0.5cm
\end{figure}

The global datasets provided by Run 2 will improve both in precision and kinematic range on previous data. Methodological uncertainties in PDF fitting have been shown to be under control, thanks to the closure test. Thus increasingly the uncertainty for which we really have no reliable estimate is the theoretical uncertainty. 

There are two categories of theoretical uncertainty. The first are the parametric uncertainties: uncertainties due to the assumed values of $\alpha_s$, $m_c$, $m_b$, $m_t$, CKM parameters, $\theta_W$, etc. Of these by far the most important is $\alpha_s$, and for this we can do no better than take the advice of the PDG. The same holds true for electroweak parameters. More controversial are the quark masses, particularly the charm mass. Attempts to determine the charm mass from the global fit itself \cite{ref:mc} are complicated by the low scale and related issues of higher twists and intrinsic charm.

The second category of theoretical uncertainty is that of missing higher order corrections. Traditionally when computing a specific cross-section these are estimated by scale variation. This method has well known failings, and is heuristic at best, but in the context of a global fit one is also faced with the issue of correlations: should the scale variations in all processes be independent, or should renormalization scales by varied together, and only factorization scales varied independently? Moreover, should factorization scales for particular types of process, for example DIS, or Drell-Yan, or jets, be treated as correlated? 

Alternative methods of estimating higher order corrections using Bayesian methods have been developed recently \cite{ref:CH}, and may be applicable to the estimation of theoretical uncertainties in PDFs. Meanwhile we can compare NLO and NNLO fits in order to estimate theoretical uncertainties: this seems to indicate that in a NLO fit the uncertainty due to missing NNLO corrections is roughly the same size as the uncertainty from the experimental data (see Fig.\ref{fig:th}), while in a NNLO fit the theoretical uncertainty is much smaller. However this could only be confirmed by performing an approximate N$^3$LO fit, perhaps based on estimates of N$^3$LO evolution and coefficient functions based on resummation.

\vskip2cm
\noindent{\bf\large Acknowledgments}

I would like to thank the organisers for inviting me to talk at this meeting, the CERN Theory Division for their hospitality while the talk was being prepared and written up, and Joey Huston for comments.


\clearpage

\begin{thebibliography}{0}
\bibitem{ref:NNPDF3}\BY{Ball R.D. et al}
\IN{JHEP}{1504}{2015}{040} [arXiv:1410:8849]
\bibitem{ref:MMHT14}\BY{Harland-Lang L.A. et al}
\IN{EPJ}{C75}{2015}{5,204} [arXiv:1412:3989]
\bibitem{ref:CT14}\BY{Dulat S. et al} [arXiv:1506.07443]
\bibitem{ref:NNPDF2}\BY{Ball R.D. et al}
\IN{Nucl.~Phys.}{B838}{2010}{136} [arXiv:1002:0744];
\SAME{B849}{2011}{296} [arXiv:1101:1300];
\SAME{B855}{2012}{153} [arXiv:1107.2652];
\SAME{B867}{2013}{244} [arXiv:1207:1303]
\bibitem{ref:MSTW08}\BY{Martin A.D. et al} \IN{EPJ}{C63}{2009}{189}
[arXiv:0901:0002]
\bibitem{ref:CT10}\BY{Lai H.-L. et al} \IN{Phys. Rev.}{D82}{2010}{074024} [arXiv:1007.2241];
\BY{Gao J. et al} \IN{Phys. Rev.}{D89}{2014}{033009} [arXiv:1302.6246]
\bibitem{ref:ABM}\BY{Alekhin S. et al} \IN{Phys. Rev.}{D81}{2010}{014032} [arXiv:0908.2766]; \SAME{D86}{2012}{054009} [arXiv:1202.2281]; \SAME{D89}{2014}{054028} [arXiv:1310.3059];
\SAME{D91}{2015}{094002} [arXiv:1404.6469]
\bibitem{ref:HERA1}\BY{Aaron F.D. et al} \IN{JHEP}{1001}{2010}{109} [arXiv:0911.0814];\BY{Abramowicz H. et al} \IN{EPJ}{C73}{2013}{2311} [arXiv:1211.1182]
\bibitem{ref:HERA2}\BY{Abramowicz H. et al} [arXiv:1506.06042]
\bibitem{ref:bench}\BY{Ball R.D. et al} \IN{JHEP}{1304}{2013}{125} [arXiv:1211.5142]
\bibitem{ref:PDF4LHC}\BY{Rojo J. et al} \IN{J.Phys.}{G42}{2015}{103103} [arXiv:1507.00556]
\bibitem{ref:norm}\BY{Ball R.D. et al} \IN{JHEP}{1005}{2010}{075} [arXiv:0912:2276]
\bibitem{ref:CTIC}\BY{Schmidt C. et al} \IN{PoS}{DIS2014}{2014}{146} 
\bibitem{ref:LHbench}\BY{Binoth T. et al} [arXiv:1003.1241]
\bibitem{ref:NNPDFth} \BY{Ball R.D. et al} \IN{Phys. Lett.}{B723}{2013}{330} [arXiv:1303.1189]
\bibitem{ref:NNPDFas}\BY{Ball R.D. et al} \IN{Phys. Lett.}{B707}{2012}{66} [arXiv:1110.2483]
\bibitem{ref:MMHTas} \BY{Harland-Lang L.A. et al} \IN{Eur.Phys.J.}{C75}{2015}{9,435} [arXiv:1506.05682]
\bibitem{ref:NNLOtop}\BY{Czakon M. et al} \IN{Phys. Rev. Lett.}{110}{2013}{252004} [arXiv:1303.6254]
\bibitem{ref:NNLOjet}\BY{Currie J. et al} \IN{JHEP}{1401}{2014}{110} [arXiv:1310.3993]
\bibitem{ref:FastNLO}\BY{Kluge T. et al} [hep-ph/0609285]
\bibitem{ref:APPLGRID}\BY{Carli T. et al} \IN{EPJ}{C66}{2010}{503} [arXiv:0911.2985]
\bibitem{ref:APFEL}\BY{Bertone V. et al} \IN{Com. Phys. Com.}{185}{2014}{1647} [arXiv:1310.1394]
\bibitem{ref:HERAfitter}\BY{Alekhin S. et al} \IN{Eur.Phys.J.}{C75}{2015}{7,304}[arXiv:1410.4412]
\bibitem{ref:RW}\BY{Ball R.D. et al} \IN{Nucl. Phys.}{B849}{2011}{112} [arXiv:1012.0836]; \SAME{B855}{2012}{608} [arXiv:1108.1758]
\bibitem{ref:GK}\BY{Giele W.T. and Keller S.} \IN{Phys. Rev.}{D58}{1998}{094023} [hep-ph/9803393]
\bibitem{ref:pumplin}\BY{Pumplin J.} \IN{Phys. Rev.}{D82}{2010}{114020} [arXiv:0909.5176]
\bibitem{ref:CJ12} \BY{Owens J.F. et al} \IN{Phys. Rev.}{D87}{2013}{094012} [arXiv:1212.1702]
\bibitem{ref:strange} \BY{Alekhin S. et al} \IN{Phys.Rev.}{D91}{2015}{9,094002}[arXiv:1404.6469]
\bibitem{ref:envelope} \BY{Botje M. et al} [arXiv:1101.0538]
\bibitem{ref:reps} \BY{Watt G. and Thorne R.S.} \IN{JHEP}{1208}{2012}{052} [arXiv:1205.4024]
\bibitem{ref:compression} \BY{Carrazza S. et al} \IN{Eur.Phys.J.}{C75}{2015}{474} [arXiv:1504.06469]
\bibitem{ref:metapdf} \BY{Gao J. and Nadolsky P.} \IN{JHEP}{1407}{2014}{035} [arXiv:1401.0013]
\bibitem{ref:MC2H} \BY{Carrazza S. et al} \IN{Eur.Phys.J.}{C75}{2015}{8,369} [arXiv:1505.06736]
\bibitem{ref:NNPDFQED} \BY{Ball R.D. et al} \IN{Nucl. Phys.}{B877}{2013}{290} [arXiv:1308.0598]
\bibitem{ref:NNPDFres} \BY{Bonvini M. et al} \IN{JHEP}{1509}{2015}{191} [arXiv:1507.01006]
\bibitem{ref:nCTEQ} \BY{Kovarik K. et al} \IN{PoS}{DIS2013}{2013}{274} [arXiv:1307.3454]
\bibitem{ref:NNPDFpol} \BY{Nocera E.R. et al} \IN{Nucl. Phys.}{B874}{2013}{36} [arXiv:1303.7236]; \SAME{B887}{2014}{276} [arXiv:1406.5539]
\bibitem{ref:mc} \BY{Gao J. et al} \IN{Eur. Phys. J.}{C73}{2013}{8,2541} [arXiv:1304.3494]; \BY{Alekhin S. et al} \IN{Phys. Lett.}{B720}{2013}{172} [arXiv:1212.2355]
\bibitem{ref:CH} \BY{Cacciari M. and Houdeau N.} \IN{JHEP}{1109}{2011}{039} [arXiv:1105.5152];  
\BY{Bagnashi E. et al} \IN{JHEP}{1502}{2015}{133} [arXiv:1409.5036]
\end{thebibliography}
\end{document}